# Rapid and accurate detection of Cyanobacterial toxin *microcystin-LR* using fiber-optic long-period grating based immunosensor

**Saurabh Mani Tripathi[a,b,*], Wojtek J. Bock[a], Predrag Mikulic[a], and Garima Mishra[c]**

[a] *Centre de Recherche en Photonique, Département d'informatique et d'ingénierie, Université du Québec en Outaouais, Gatineau, QC, J8Y 3G5, Canada*
[b] *Department of Physics, Indian Institute of Technology Kanpur, 208016, India*
[c] *Department of Structural Biology, Weizmann Institute of Science, Rehovot, Israel*
*[*smt@iitk.ac.in](mailto:smt@iitk.ac.in)*

**Abstract:** In this paper we present a label-free, stable and highly sensitive fiber-optic sensor for detection of environmental toxin microcystin-LR (MC-LR). Thiolated MC-LR-targeting aptamer covalently immobilized on dual-resonance long-period fiber gratings (DR-LPFG) is used as MC-LR recognition element and the spectral response of the DR-LPFG is monitored for various concentrations of MC-LR. Sensitivity of the DR-LPFG is optimized by coating an appropriate thickness of gold layer to the fiber surface. Gold coating serves two purposes, first it locally supports a surface-Plasmon mode associated with the modified cladding modes and second it helps covalently immobilize the MC-LR targeting aptamer to the sensor surface. Monitoring both the resonance wavelengths of our ultra sensitive DR-LPFGs (3891.5 nm/RIU) we study the aptamer and MC-LR binding to the sensor surface. Real-time, efficient MC-LR molecular binding to the sensor surface is also confirmed using atomic force microscopy. The detection limit of our sensor, limited by the molecule size of MC-LR, its fractional surface coverage of the sensor, and sensor length, is 5 ng/mL, which can be further improved by increasing the fractional MC-LR coverage on the sensor surface by (*i*) reducing the LPFG length (*ii*) increasing grating strength, (*iii*) increasing the LPFG sensitivity by coupling to even higher order cladding modes by reducing the grating period.

## 1. Introduction

The cyclic heptapeptides microcystins (MCs) produced by several bloom-forming cyanobacteria, e.g. Anabaena Microcystis, and Planktothrix, are one of the most dangerous and widely studied microcystins (Messineo et al, 2009; Sivonen et al, 2009). Among these, the microcystin-LR (MC-LR) is the most frequently present and toxic MC. Being hepatotoxic in nature, the MC-LR primary targets the liver and promotes the formation of hepatic tumors by chronic ingestion through contaminated food or water. Consumption of a large dose can be deadly among all age groups as the MC-LR can also affect the kidney and the gastrointestinal tract (Azevedo et al., 2002; WHO). The symptoms generally include headache, blurred vision, abdominal pain, nausea, and vomiting etc. The MC-LR is released in substantial amounts during the bacterial cell lysis. Being chemically stabile and highly soluble to water, it can maintain its presence for a long time in the surface water bodies. Apart from their exposure through water, the MC-LR also poses a great risk of exposed through the consumption of freshwater fish, seafood and micro-algae dietary supplements (Ibelings et al., 2007; Ortelli et al., 2008). With the incidence of harmful cyanobacterial algal blooms becoming more frequent and the extent of coverage increasing worldwide, the need for their routine and accurate remote detection is the primary challenge for drinking and recreational water activities. The median lethal dose ($LD_{50}$) of the MC-LR is 5 mg/kg and the World Health Organization (WHO) has established the maximum concentration limit for the microcystins in daily drinking water at 1 μg/L (WHO/SDE/WSH/03.04/57).

For their detection the currently used technologies rely mainly on two methods: the analytical methods, like reversed-phase liquid chromatography (LC), or the biochemical screening methods such as enzyme-linked immunosorbent assays (ELISA) (Lindner et al, 2004; Pyo et al, 2005; Mountfort et al., 2005; Sangolkar et al., 2006), competitive enzyme immunoassays (Khreich et al., 2009; Long et al., 2009), or protein phosphatase inhibition assays (PPIA) (Mountfort et al., 2005; Allum et al., 2008). Both of these methods, however, have their inherent disadvantages. For example, the LC method often requires time consuming sample preparation procedures involving the pre-concentration of considerable water volumes prior to LC analysis. The ELISA and PPIA, on the other hand, lack from distinguishing between various microcystins and other unrelated protein phosphatase inhibitors (Sangolkar et al., 2006). Recently, various immunosensors based on electrochemical (Loyprasert et al., 2008), nuclear magnetic resonance (Ma et al., 2009), and various optical measurements (Hu et al., 2009; Lindner et al., 2009; Long et al., 2009) have also been reported. Although offering high sensitivity, these methods often require skilled operators, pre-treatment, and expensive instrumentation, making them unsuitable for routine pollution monitoring. Furthermore, none of these techniques is suitable for remote monitoring, making it necessary to transport the bio-sample to the distant laboratory rather than deploying the sensor at the point of need itself.

In order to obviate these difficulties, the fiber optical methods seem to be a natural choice. They are compact in size, easy to use, inexpensive, support remote sensing, offer the possibility of multiplexing many sensors to a single system, and often do not require any specialized training for their operation. The major difficulty associated with these sensors, however, is their low sensitivity: ~ 1000 - 1500 nm/RIU (RIU refractive index unit) for bio-samples (refractive-indices 1.33-1.34), offered by the best optical sensors including the surface Plasmon polaritons (SPP) (Homola, 2006), long-period gratings (LPGs) (Kashyap, 2009; Tripathi et al., 2008) etc., which is insufficient for an accurate detection of small molecules like MC-LR. Further, binding the MC-LR molecules to the optical fiber surface is also quite challenging.

In this paper we address these issues and present the development of a fiber-optic MC-LR immunosensor based on the covalent immobilization of MC-LR targeting aptamer on extremely sensitive dual-resonance long-period fiber-optic gratings (DR-LPFGs). The DR-LPFGs are a specific class of long-period gratings which excite two resonance wavelengths corresponding to the same cladding mode and grating period.

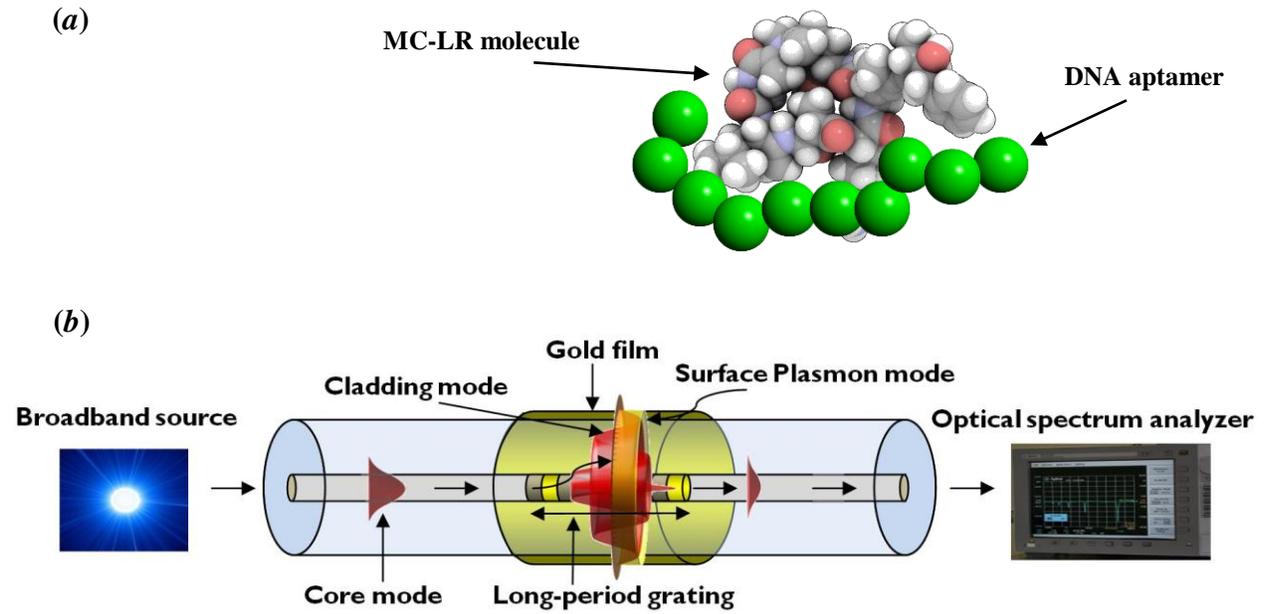

Fig.1. Schematic representation of (*a*) MC-LR binding by DNA aptamer, and (*b*) the sensor structure used in our experiments.

The resonance wavelength ($\lambda_R$) is can be mathematically expressed as (Kashyap, 2010),

$$\lambda_R = \Lambda \left( n_{eff}^c - n_{eff}^{cl} + \frac{\kappa_{c-c} - \kappa_{cl-cl}}{k_0} \right) \quad (1)$$

where, $\Lambda$ is the grating period; $n_{eff}^c$ and $n_{eff}^{cl}$ are the effective refractive indices of the core mode and cladding mode, respectively; $\kappa_{c-c}$ and $\kappa_{cl-cl}$ are the self-coupling coefficients of the core mode and cladding mode, respectively, and $k_0$ is the free space propagation constant. The operating principle of LPFG based bio-sensors is as follows. The evanescent field associated with the cladding mode interacts with the ambient region, changing the $n_{eff}^{cl}$ and thus changing the $\lambda_R$. By measuring the changes in $\lambda_R$ as a function of varying ambient region perturbations, the changes in the ambient perturbation are quantified. An additional contribution to the changes in the $\lambda_R$ comes from the DR-LPFG; the two $\lambda_R$ shift in opposite directions for the DR-LPFG. Therefore, we can define the sensitivity of a DR-LPFG sensor as the fractional change in the separation of two resonance wavelength with respect to the external perturbation. Thus, for our MC-LR immunosensor, we define the sensitivity as $S = \frac{\Delta(\lambda_{R2} - \lambda_{R1})}{\Delta MC - LR}$, where ΔM-LR is the change in the MC-LR concentration. The schematic diagram of the sensor structure used in our experiments is shown in Fig.1, where we have plotted (*a*) MC-LR molecular binding to the aptamer and (*b*) the LPFG sensor detection mechanism; note a local-surface Plasmon excited at the metal/dielectric boundary. In the following we first discuss the experimental development of the DR-LPFG sensor and then present our experiment results and discussion.

## 2. Materials and methods

### 2.1 Reagents

Microcystin-LR was purchased from Enzo Life Sciences, USA, and was dissolved into 50 mM Tris, pH 7.5, (150mM NaCl, 2mM $MgCl_2$) binding buffer. This stock was further diluted to the required concentrations during the experiments. The 5'- disulphide terminated MC-LR-targeting aptamer was chemically synthesized and purified by PAGE (IDT, USA) (Andy Ng et al.; 2012), allowing attachment to gold surface via the Sulphur atom. The aptamer was dissolved in buffer at 100 μM as the stock solution, and diluted further to the required concentrations during the experiment.

### 2.2 Sensor fabrication and resonance wavelength tuning

We fabricated several LPFGs in hydrogen loaded (150 bar, for 15 days) telecommunication-grade single mode optical fiber (Corning: SMF-28™) using a chromium amplitude mask (Λ= 226.8 μm) and high-power KrF excimer laser (Lumonics™ Lasers: Pulse Master®-840) emitting at 248 nm, at a pulse repetition rate of 100 Hz, pulse duration of 12 ns and peak pulse energy of 10 mJ (Tripathi et al., 2012). The LPFGs were then thermally annealed at 150 °C for 3 hours to release excess hydrogen and stabilize the grating's optical properties. Finally the cladding region was partially etched in 4% HF for ~2.3 hours to tune the resonance wavelength close to the turn-around wavelength. In Fig.2 we have plotted the experimental transmission spectrum at the beginning ($\Delta\tau_{etch} = 0$, dashed curve) and end ($\Delta\tau_{etch} = 138'\ 20''$, solid curve) of etching process; the evolution of resonance wavelength is also shown in the inset. The turnaround wavelength at ~1595 nm is evident from the inset and the dual resonance behavior can be easily observed from the solid curve in Fig2. Owing to their operation in dual-resonance regime, these LPFGs are extremely sensitive to any changes in their cladding diameter (dw) as well as the ambient refractive index ($dn_{se}$). The typical surface sensitivity and refractive-index sensitivity for our gratings are $d(\Delta\lambda_R)/dw = 1.3$ nm/nm, and $d(\Delta\lambda_R)/dn_{se} = 3096.5$ nm/RIU, respectively; where $\Delta\lambda_R = \Delta(\lambda_{R2}-\lambda_{R1})$ is the net change in the separation between two resonance wavelengths.

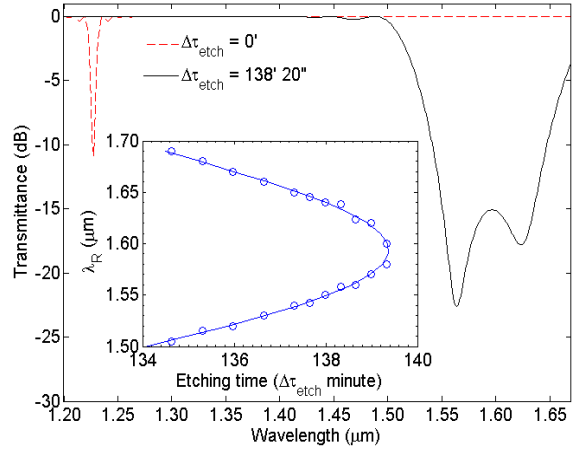

Fig.2. Transmission spectrum evolution with respect to partial cladding etching; the resonance wavelength tuning has been shown in the inset.

*2.3 Metal deposition and sensitivity optimization*

Next, to facilitate the aptamer immobilization on LPFG sensor we deposited gold layer on the cylindrical surface of the fiber using thermal vapor deposition method. In our experiments we observed that the gold coating on LPFG has two consequences: (*i*) it locally supports a surface-Plasmon mode associated with the modified cladding modes (see Fig.3(*a*) and (*b*), obtained for an optical fiber with its core and cladding regions made of 4.1 mole% GeO₂ doped SiO₂ and fuzzed SiO₂, respectively, with their respective diameters being 8.2 μm and 110 μm) and thereby increases the evanescent field in the ambient region, enhancing the LPFG sensitivity, and (*ii*) gold coating also increases the cladding thickness and thereby shifts the two resonance wavelengths ($\lambda_{R1}$ and $\lambda_{R2}$) wide apart, reducing the sensitivity considerably. Thus, there is a tradeoff between these two effects and an optimum metal thickness has to be chosen to obtain the maximum sensitivity (Tripathi et al., 2008).

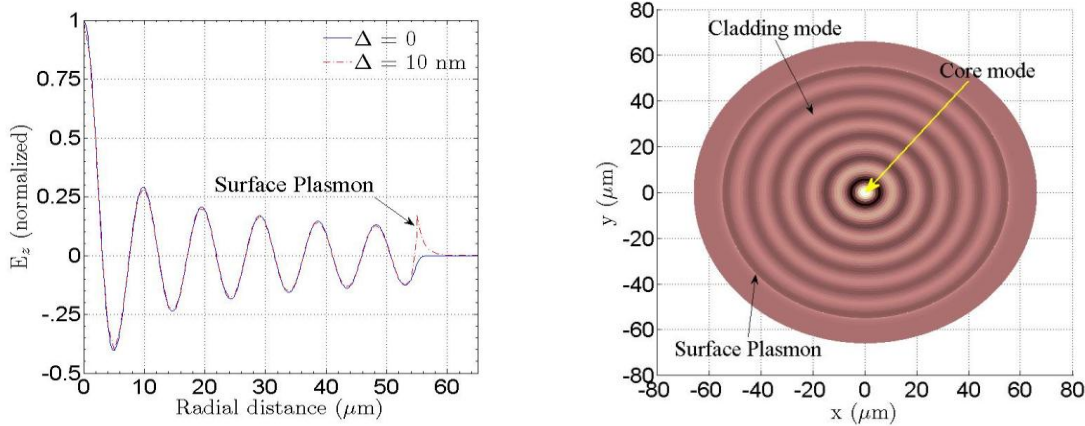

Fig.3. Theoretical normalized filed distribution of the excited cladding mode in the presence and absence of gold coating on the optical fiber. The local surface Plasmon mode supported at gold/dielectric boundary is evident as (*a*) a spike in radial field distribution and (*b*) a thin ring in azimuthal field distribution.

Due to a high dielectric constant of gold ($\varepsilon_d$ ~104) even a small thickness of gold layer shifts the resonance wavelength wide apart. For our sensor we observed a change of ~20 nm in resonance wavelengths per 1 nm change in gold thickness deposited on the fiber surface. Therefore, in order to keep the resonance wavelengths close to the turn-around wavelength even with the metalized LPFGs, we further etched the LPFGs till the two resonances merge together and eventually disappear completely. The resulting transmission spectrum for a higher ARI (= 1.38) is shown in Fig.4 (dotted curve). Now, prior to Au deposition, to increase the adhesion of Au on $SiO_2$ surface (fiber cladding) in order to avoid it's peeling-off during subsequent washes of the sensor, we first deposited a thin Cr film (~2.5 nm) on the fiber cladding using the sputtering method (Luminos Cluster R&D Tool System, Canada). This step was then followed by the thermal vapor deposition of Au layers of different thicknesses, in a range of 5 – 20 nm, on fiber surface. To ensure a uniform metal deposition the LPFGs were fixed on a rotational stage rotating at a speed of 60 rpm during deposition process. The transmission spectra recorded after the Cr and Au films deposition are shown by dash-dotted and solid lines, respectively, in Fig.4. The re-emergence of the transmission spectrum after metal depositions confirms the efficient metal deposition on the LPFGs. The ARI sensitivities with respect to different gold thicknesses were then measured using analytes of known refractive indices measured by Abbe refractometer (Atago DR-M2) operating at $\lambda = 1.55$ μm. For our gratings the highest sensitivity is obtained at a Gold thickness of ~10 nm (Tripathi et al, 2008). The variation of the two resonance wavelengths with respect to the analyte refractive indices are shown in the inset. The measured refractive index sensitivities of the two resonance wavelengths are ~1797.1 nm/RIU and ~2094.4 nm/RIU, respectively, giving the net sensitivity as 3891.5 nm/RIU; which, to the best of our knowledge, is the highest reported refractive index sensitivity of the LPFGs based sensors.

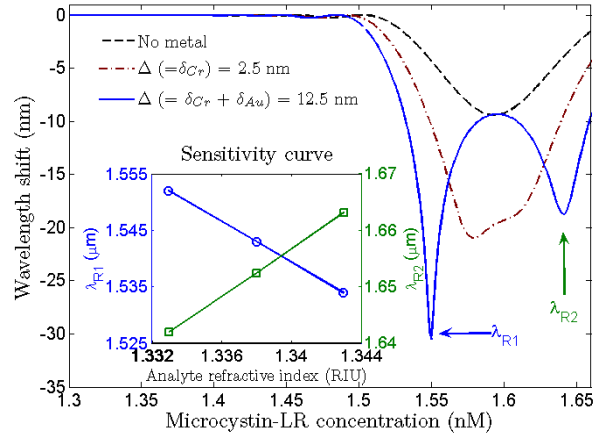

Fig.4. Transmission spectrum evolution with different metal thicknesses; the ambient region is water. The ambient refractive index sensitivity curves are shown in the inset.

Finally, we determined the quality of the metal films so deposited (on a background slide made of silica) using a broadband spectroscopic ellipsometer (HORIBA Jobin Yvon). The mean thicknesses of Cr and Au films (taking measurements on 5 different sites) are 1.52 (+/- 0.02) nm and 5.02 (+/-0.02) nm, respectively, with an air-filling fraction of ~25%. It is important to mention here that unlike the pure surface-Plasmon mode, which has its field maxima at metal/etched-cladding boundary and decays exponentially in all the regions beyond it (Nemova and Kashyap, 2006; Tripathi et al. 2008), the surface-Plasmon modes excited in our sensor are local in nature (Tripathi et al., 2009), *i.e.* the field associated with the cladding modes gets modified at the metal/cladding boundary showing a local maxima at the metal/dielectric boundary as shown in Fig.3. Therefore, their excitation and modal properties will not be affected much by the surface roughness present in our sensor.

## 3. Experimental procedure, results and discussion

For MC-LR detection experiments the LPFG sensor was passed through a flow cell, and in order to avoid any micro-bends a fixed force to the fiber was maintained throughout the experiment. Light was launched into the fiber using an *Agilent*-83437A broadband source (BBS) and the transmission spectrum was recorded using an *Agilent*-86142B optical spectrum analyzer (OSA) with a resolution of 0.02 nm. To bind the MC-LR to sensor surface we first neutralized the pH of the LPFG surface by thoroughly washing it in a buffer of pH 7.5. Next the LPFGs were incubated in 5' thiolated MC-LR-targeting aptamer (200 nM in binding buffer) for ~3 hours to allow covalent attachment of aptamer to the gold surface coated on the LPFGs. The LPFGs were then washed copiously with the binding buffer and subsequently incubated with various concentrations of MC-LR for ~60 minutes. Finally the LPFGs were washed again in the buffer. The transmission spectrum of the LPFGs was continuously recorded during various processes of the experiment.

The AFM micrographs (VEECO AFM) of the sensor surface after APTAMER immobilization and (*a*) pre-buffer wash to MC-LR incubation (*b*) after MC-LR incubation and (*c*) post-buffer wash to MC-LR incubation for two different concentrations of the MC-LR are shown in Fig. 5. The physical deformation of the metallic surface due to MC-LR binding is quite evident from these micrographs. The corresponding deflection signals are also shown in Fig.5(*d*)-(*f*).

**(*a*) Pre-buffer wash   (*b*) MC-LR incubation   (*c*) Post-buffer wash**

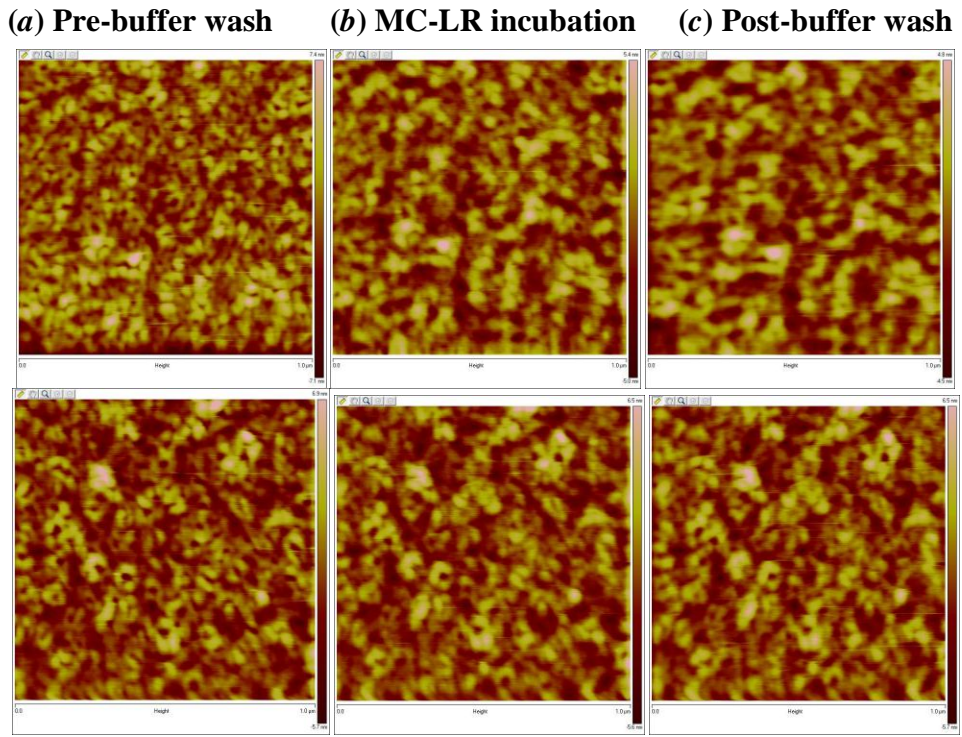

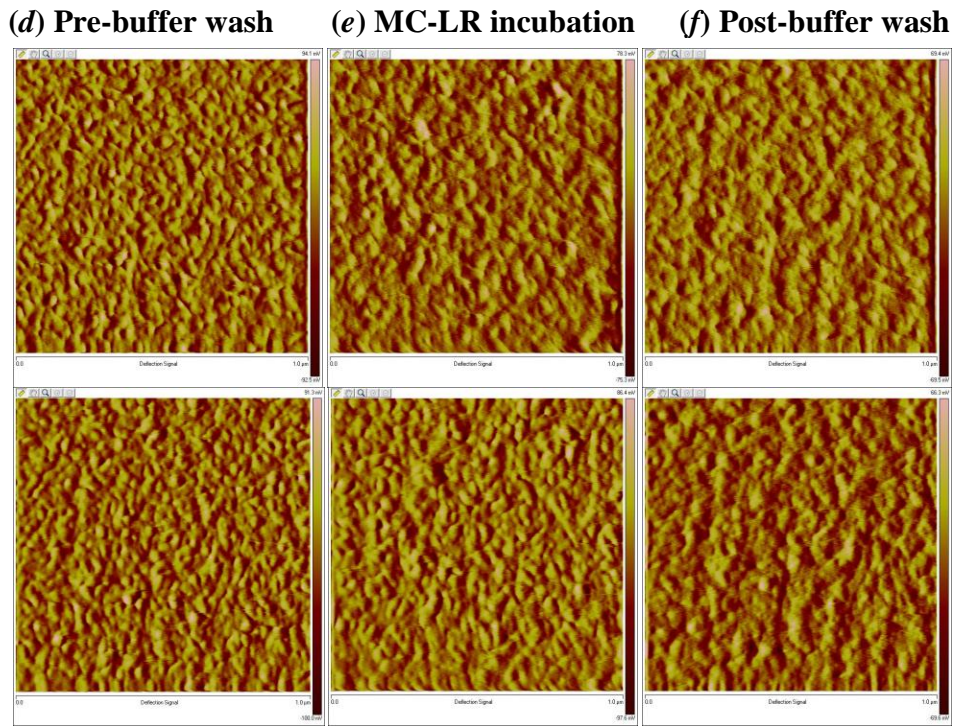

Fig.5. AFM micrographs of the sensor surface after APTAMER immobilization and (*a*) pre-buffer wash to MC-LR incubation (*b*) after MC-LR incubation and (*c*) post-buffer wash to MC-LR incubation for various concentrations of the MC-LR.

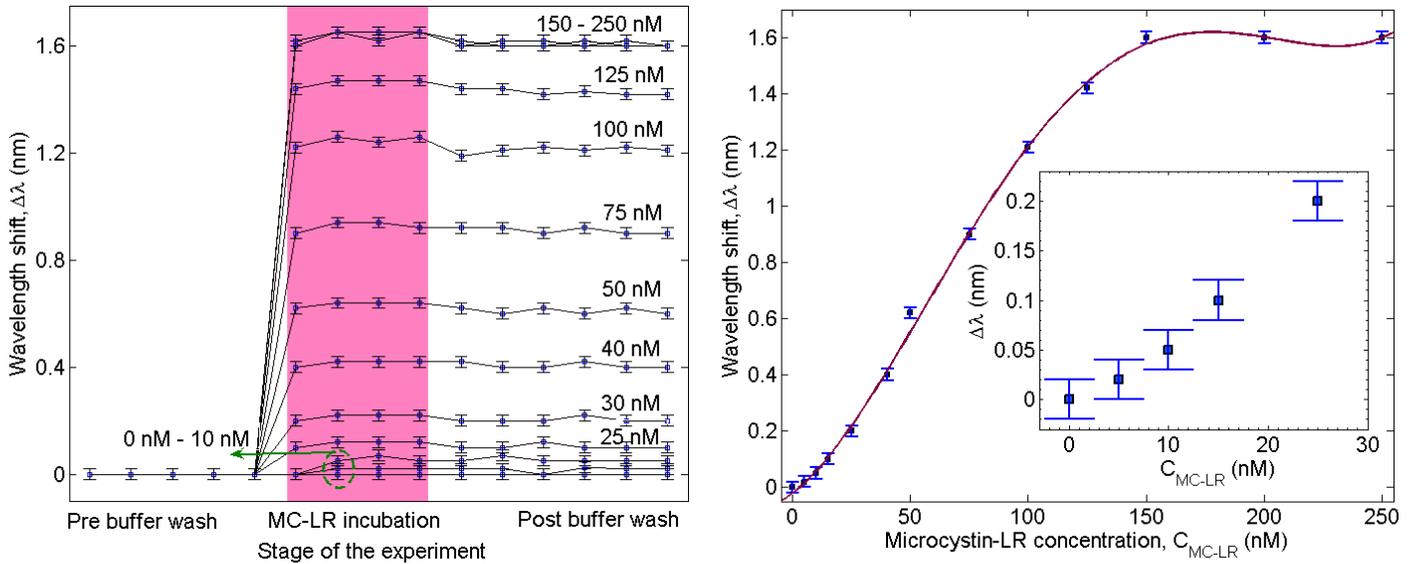

Fig.6. Variation of (a) the spectral shift of the resonance wavelength prior to and after the *MC-LR* incubation on the LPFG, and (b) the spectral shift of the resonance wavelength as a function of the *MC-LR* concentration in the buffer.

The net wavelength shifts ($\Delta\lambda_R = \Delta|\lambda_{R1} - \lambda_{R2}|$) recorded during the last three stages of our experiments, namely *pre* buffer wash – MC-LR incubation – *post* buffer wash, are shown in Fig.6(a). The overall response time of the sensor, after the MC-LR incubation, is ~ 60 minutes and the transmission spectrum can be continuously monitored using an optical spectrum analyzer. This sensor, thus, provides an attractive, reliable and very fast way to monitor the growth/decay of MC-LR culture. The lowest concentration of MC-LR measured in our experiment is 5 nM (~4.98 ng/mL; molecular weight of the MC-LR is 995.17 g.mol$^{-1}$), the corresponding resonance wavelength shift is 0.02 nm (which is also the accuracy of our measurements). Nevertheless, the detection limit can be further improved by employing higher spectral resolution detection system with typical resolution of ~ 1pm (Tripathi et al, 2008; HP manual).

It is worthwhile mentioning here that unlike the widely used refractive-index measurement based bio-sensors, the spectral shifts observed in our experiments are influenced not only by the changes in refractive index of the bio-solutions but also by the real MC-LR binding to the sensor surface. The resonance wavelength shifts in our experiments are, therefore, contributed due to both (*i*) the changes in the bio-samples' refractive-indices ($\Delta n_{ARI}$) due to various concentrations of MC-LR and (*ii*) the changes in fiber cladding radius ($\Delta r_{cl}$) due to MC-LR binding.

In order to estimate the influence of MC-LR on the refractive index variation of the bio-sample, we measured the refractive indices of various concentrations of the MC-LR (in buffer) using an Abbe refractometer (Atago DR-M2) with an accuracy of 0.0001 RIU, shown in Fig. 7. The overall refractive index difference in the buffer solution, is ~ 2×10$^{-4}$ RIU for the MC-LR concentrations varying between 0 nM to 150 nM. This implies that, based on an exclusive refractive index sensing of MC-LR, the LPFGs used in our experiments (with a sensitivity of 3891.5 nm/RIU) should yield a net resonance-wavelength shift of ~ 0.78 nm. This shift is much smaller than the ~ 1.6 nm spectral shifts observed in our experiments: a further confirmation of an efficient MC-LR binding on the optical fiber surface. Another effect quite evident form Fig. 7 is that it would be quite difficult to measure the changes in the MC-LR concentration based on the refractive-index measurement alone. For example, the used Abbe refractometer shows the same refractive index for the MC-LR concentrations in the range 0-25 nM; 50-75 nM; and 100-150 nM etc. In contrast our sensor is capable to efficiently predict the changes in MC-LR concentrations in the very small steps of 5 nM, which can be further reduced by reducing the LPFG length.

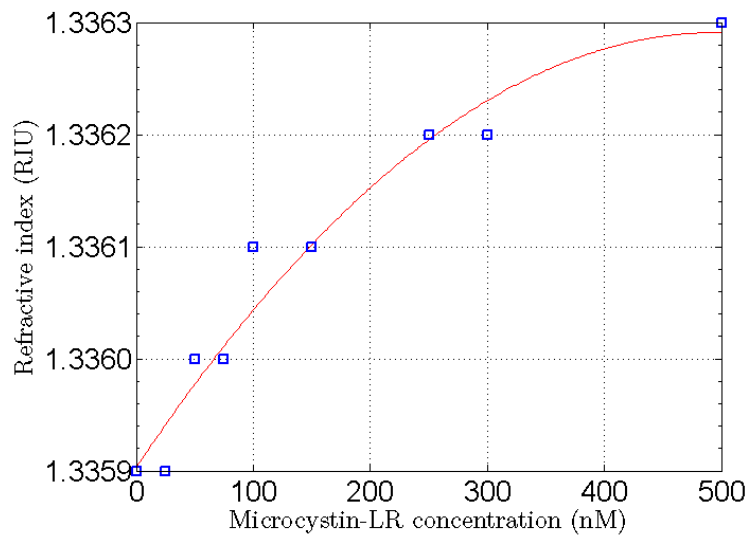

Fig.7. Variation of the refractive index (measured at $\lambda = 1.55$ μm) of the bio-sample as a function of the *MC-LR* concentration in the buffer.

## 4. Conclusions

In this paper we have presented a rapid, reliable, cost-effective, and label-free sensor for detection of the environmental toxin microcystin-LR. Using the highly accurate spectral interrogation mechanism we monitored the specific MC-LR binding to a DNA aptamer covalently immobilized on the LPFG surface. Resonance-wavelength shifts ranging between 0.02 nm to 1.6 nm for MC-LR concentrations varying between 5 nM and 250 nM have been observed; the corresponding experimental accuracy varying between 50% - 99.38%. Due to the covalent binding between the DNA aptamer and the gold coated LPFG surface, no MC-LR dissociate during subsequent washes of the sensor (as they do in physical adsorption-based measurement procedures), making the measurements highly stable compared to those of adsorption-based sensors. We showed that the present sensor can efficiently detect the MC-LR concentration as low as 5 nM, which can be further lowered by increasing the fractional MC-LR coverage on the sensor surface by (*i*) reducing the LPFG length (*ii*) increasing grating strength, (*iii*) increasing the LPFG sensitivity by coupling to even higher order cladding modes by reducing the grating period.

**Acknowledgments**

The present work was partially supported by the Natural Science and Engineering Research Council of Canada; by Canada Research Chairs program; and by INSPIRE grant (grant number IFA13-PH-69) of the DST, Ministry of Science & Technology, Govt. of India.